\pdfoutput=1

\documentclass{article}

\usepackage{amsmath,amsthm}
\usepackage{booktabs} %
\usepackage[utf8]{inputenc}
\usepackage{hyperref}
\usepackage{boxedminipage}
\usepackage{csvsimple}
\usepackage{cleveref}
\usepackage{xspace}
\usepackage{multirow,multicol}
\usepackage{xargs}
\usepackage{pgfplots}
\usepackage{pgf-umlsd}
\usepgflibrary{arrows}
\usepackage{tikz}
\usetikzlibrary{shapes,arrows,chains}
\usepackage[round-mode=places, round-integer-to-decimal, round-precision=2,
    table-format = 1.2,
    table-number-alignment=center,
    round-integer-to-decimal]{siunitx}
\usepackage{authblk}
\usepackage[newfloat,frozencache]{minted}
\setminted[]{breaklines=true,
             fontsize=\small}

\newcommand{\minorsubsection}[1]{\textbf{#1}\xspace}

\newcommand{\toolname}[1]{\emph{#1}}

\newcommand{\cmpToUs}{\minorsubsection{Comparison to our approach}}

\newcommand{\inlineJS}[1]{\mintinline{JavaScript}{#1}}
\newcommand{\Unexpected}{Unexpected behavior\xspace}
\newcommand{\unexpected}{unexpected behavior\xspace}
\newcommand{\Ie}{I.e.,\xspace}
\newcommand{\ie}{i.e.,\xspace}

\newcommand{\eg}{e.g.,\xspace}

\newcommand{\ourtool}{\toolname{SAFE-PDF}\xspace}

\newcommand{\CC}{\toolname{Clean Content}\xspace}

\usepackage{pifont}%
\newcommand{\cmark}{\ding{51}}%
\newcommand{\xmark}{\ding{55}}%

\newcommand{\authnote}[2]{{\bf \textcolor{blue}{#1}: \em \textcolor{red}{#2}}}

\renewcommand{\authnote}[2]{}

\usepackage{etoolbox}
\BeforeBeginEnvironment{tabular}{\begin{center}\footnotesize}
\AfterEndEnvironment{tabular}{\end{center}}

\author[1]{Alexander Jordan}
\author[1]{Fran\c{c}ois Gauthier}
\author[1]{Behnaz Hassanshahi}
\author[1,2]{David Zhao}
\affil[1]{Oracle Labs, Brisbane, Australia}
\affil[ ]{\{\textit{firstname.lastname}\}@oracle.com}
\affil[2]{University of Sydney, Sydney, Australia}
\affil[ ]{d-z@outlook.com}

\title{SAFE-PDF: Robust Detection of JavaScript PDF Malware With Abstract Interpretation}

\begin{document}
\maketitle

\begin{abstract}
The popularity of the PDF format and the rich JavaScript environment that PDF
viewers offer make PDF documents an attractive attack vector for malware
developers. PDF documents present a serious threat to the security of 
organizations because most users are unsuspecting of them and thus likely
to open documents from untrusted sources.

We propose to identify malicious PDFs by using conservative abstract interpretation to
statically reason about the behavior of the embedded JavaScript code.
Currently, state-of-the-art tools either: (1) statically identify PDF malware
based on structural similarity to known malicious samples; or (2) dynamically
execute the code to detect malicious behavior. These two approaches
are subject to evasion attacks that mimic the structure of benign documents or
do not exhibit their malicious behavior when being analyzed dynamically. In contrast, abstract 
interpretation is oblivious to both types of evasions.
A comparison with two state-of-the-art PDF malware detection tools shows
that our conservative abstract interpretation approach achieves similar
accuracy, while being more resilient to evasion attacks.

\end{abstract}

\section{Introduction}
The Portable Document Format (PDF) allows for the embedding of interactive
elements written in JavaScript\footnote{%
While the major parts of the PDF format are standardized as ISO 32000-1~\cite{ISO32000},
the specification of some advanced features supported by Adobe's PDF software
remains proprietary technology, and is referenced only by the ISO standard.
The proprietary parts include JavaScript support and APIs available in Adobe's
PDF software. Adobe provides an informal specification for this JavaScript part
in their public documentation~\cite{AdobePDF3DAPI, AdobePDFAPI}.}.
JavaScript in PDFs allows document creators to support input validation in 
forms and to offer convenient shortcuts for common actions such as printing.
More elaborate use cases of PDF JavaScript include controlling embedded
multimedia objects and interacting with the file system
or network. However, this rich and complex PDF JavaScript environment can also 
be used for illegitimate purposes. Indeed, previous work has shown that JavaScript
is the vector of choice for PDF malware because: (1) implementation bugs in the 
PDF JavaScript extensions can be exploited to deliver and execute malicious
payloads; (2) bugs in the JavaScript runtime and/or sandbox can be triggered
with JavaScript code; and (3) JavaScript can be used as a facilitator to 
exploit vulnerabilities \emph{outside} the JavaScript environment through 
techniques like heap spraying~\cite{Laskov2011,carmony2016}.

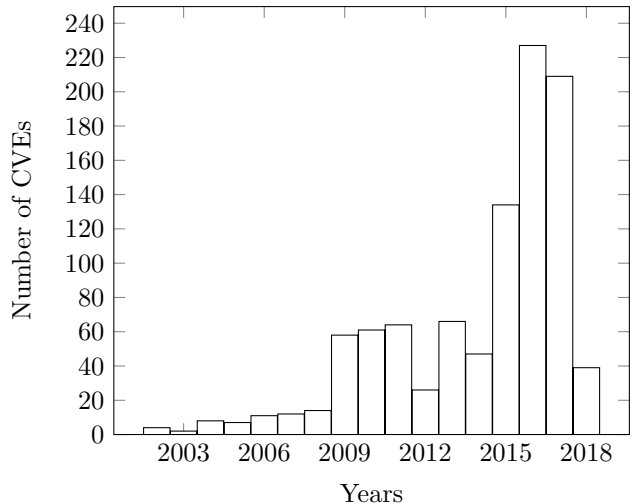
\begin{figure}[t]
\centering
\begin{tikzpicture}
  \begin{axis}[
    symbolic x coords={2002, 2003, 2004, 2005, 2006, 2007, 2008, 2009, 2010, 2011, 2012, 2013, 2014, 2015, 2016, 2017, 2018},
    ylabel = {Number of CVEs},
    xlabel = {Years},
    xtick={2003, 2006, 2009, 2012, 2015, 2018},
    ytick distance=20,
    ymin=0]
    \addplot[ybar,fill=white] coordinates {
      (2002,4)
      (2003,2)
      (2004,8)
      (2005,7)
      (2006,11)
      (2007,12)
      (2008,14)
      (2009,58)
      (2010,61)
      (2011,64)
      (2012,26)
      (2013,66)
      (2014,47)
      (2015,134)
      (2016,227)
      (2017,209)
      (2018,39)
    };
  \end{axis}
\end{tikzpicture}
\caption{Number of CVEs containing ``Adobe Reader'' in their description.}
\label{fig:cve}
\end{figure}

In 2008, the number of PDF-based attacks increased sharply. Then, in 2009,
the number of Common Vulnerabilities and Exposures (CVEs) reported against
the Adobe Reader rose alarmingly. Previous work showed how the vast majority of 
PDF malware uses JavaScript in one way or another~\cite{risepdfmalware,Laskov2011,carmony2016}. 
The histogram in \Cref{fig:cve} shows how the number of CVEs reported against
Adobe Reader is still at all time highs, with 39 CVEs already reported at the time
of writing (May 2018), despite the introduction of sandboxing for increased
safety in Adobe Reader X.

To lower the risk to users posed by PDF malware, several well-known
techniques are available.
These techniques are either employed in the PDF viewer software
or implemented as (part of) stand-alone software products, such as anti-virus
or anti-malware scanning tools.

We found, however, that existing tools to detect PDF malware (both in industry and research),
are vulnerable to various kinds of evasion attacks. %
Indeed, simple static malware detectors (e.g. anti-virus tools)  
that rely on a signature database that contains known patterns of exploit code are
subject to code obfuscation attacks. More recent and advanced approaches that use 
machine learning to automatically capture structural patterns of malware are
vulnerable to mimicry attacks where the malware emulates the structure of benign 
documents to avoid detection. Finally, approaches that attempt to statically isolate and 
analyze PDF JavaScript code are typically vulnerable to parser confusion attacks,
where the malware is embedded in non-standard, or poorly documented PDF constructs
that are meant to trip up code extractors. 

\emph{Dynamic} (runtime) malware detection tools, on the other hand, rely on executing 
code under analysis in a special environment (a \emph{sandbox}) to detect suspicious
behaviors (e.g., network traffic, execution of known exploit code).
However, dynamic approaches are susceptible to \emph{sandbox evasion} attacks where the 
malicious code avoids detection by probing its environment to detect if it is running in 
a sandbox environment~\cite{Chen2008}. Dynamic approaches also typically load suspicious
PDFs in a limited number of viewer applications, which makes them fail to identify malware that
checks the version of the viewer before picking a vulnerability to exploit. More importantly,
dynamic approaches are under-approximative by nature, \eg they may miss any malicious
behavior that does not manifest itself by loading the document only, but
is rather triggered by user interaction.

To address the limitations of signature-based, machine learning, and dynamic tools, we 
introduce \ourtool, an abstract interpretation-based JavaScript malware detector for PDF 
documents. Abstract interpretation is a powerful framework for static program analysis that
computes a sound over-approximation of all possible program behaviors. In other words,
abstract interpretation allows us to check if a JavaScript program can \emph{ever} 
exhibit malicious behavior under \emph{any} possible execution and viewer version.
Furthermore, we address the main limitation of existing static analysis approaches by
developing one of the most robust PDF JavaScript code extractors to date and validating it
against 2952 documents that are known to cause issues in existing code extractors~\cite{carmony2016}. 
We also complement abstract interpretation with a model of the JavaScript
environment available to programs embedded in a PDF document, and a whitelist policy
that does not allow malicious behavior (\eg calls to vulnerable APIs, heap
spraying) to go undetected. The model and whitelist are easy to understand and 
maintain, especially when compared to opaque machine learning classifiers.
Used together with our analysis in a \emph{conservative} manner, obfuscation, mimicry,
and sandbox evasion attacks are no longer effective.

This prompted us to design the \ourtool tool for high recall%
\footnote{ratio of true positives to total number of malware in a given set of documents}.
and robustness to evasion attacks.
At the same time, \ourtool (due to its straight-forward model and whitelist)
is flexible enough to be quickly adapted to
changes in the threat landscape, because it does not require training or tuning
of parameters.
In this paper, we show how \ourtool
achieves comparable precision%
\footnote{ratio of true positives to total number of reports}
and recall to state-of-the-art tools, while being oblivious
to evasion strategies that affect existing approaches.
To summarize, we make the following contributions:
\begin{itemize}
  \item We present a practical conservative abstract interpretation approach that
    efficiently detects PDF malware by statically analyzing \emph{all possible behavior} 
    of its embedded JavaScript code in less than 4 seconds on average.
  \item We develop one of the most robust PDF JavaScript code extractors to date, by
    addressing all known limitations of existing extractors and testing our tool on PDF 
    documents with hard-to-extract JavaScript code~\cite{carmony2016}.
  \item We compare \ourtool against two state-of-the-art PDF malware
    detection tools: \toolname{PDF Malware Slayer}~\cite{Maiorca2012}, 
    and \toolname{Hidost}~\cite{Vsrndic2016}, and show how \ourtool
    achieves comparable precision and recall while capturing malware that evade other
    tools.
\end{itemize}
\section{Overview}

\subsection{Motivation}

PDF malware frequently depends on embedded JavaScript code.
While disabling JavaScript in the PDF viewer and removing it 
from documents is the most obvious and effective defense against JavaScript PDF malware, we
think that such measures impose an unnecessary loss of functionality on users. Indeed, blind removal 
of JavaScript code breaks several harmless PDF documents and prevents users from using very 
common and useful features such as form input validation. \ourtool thus aims at
enabling JavaScript use in benign documents while \emph{conservatively} and \emph{robustly} 
filtering out malicious documents.

JavaScript malware in PDF
documents seeks to exploit bugs in PDF viewer applications.
As a result, an attacker is able to disrupt operation or gain control of the targeted host system.
The most frequently exploited vulnerabilities in PDF viewer applications have shown to be:
(1) lack of user input validation and the allowing of arbitrary and unsafe operations 
outside the scope of the document;
(2) unintended side effects of otherwise legitimate operations (e.g. file creation, or network 
access);
and (3) memory corruption bugs in PDF JavaScript extensions or in the runtime itself.

\subsection{Existing approaches to PDF malware detection}

\begin{listing}
\begin{minted}{JavaScript}
try {
    eval("thi" + "s.f" + "ea" + "tu" + "re" +"A.n" + "ew" + "Ob" + "je" + "ct" + "(n" + "ull)");
}
\end{minted}
    \caption{Artificial malware example: evaluating string with simple obfuscation of call to \mintinline[fontsize=\small]{JavaScript}{featureA.newObject(size)}}
    \label{lst:obf1}
\end{listing}

\begin{listing}
\begin{minted}{JavaScript}
function start(foo, bar) {
    // bar = reference to featureB object
    // encoded property = "sendMessage"
    bar["\x73\x65" + "\x6e" + "\x64\x4d" + "\x65\x73" + "\x73\x61" + "\x67\x65"]({
        data: /* ... */
    });
}
\end{minted}
    \caption{Artificial malware example: obfuscated call to exploitable function \mintinline{JavaScript}{featureB.sendMessage()}}
    \label{lst:obf2}
\end{listing}

\begin{listing}
\begin{minted}{JavaScript}
function urpl(sc) {
    var keyu = "%u";
    var re = /XY/g;
    sc = sc.replace(re, keyu);
    return sc;
}
var unes = unescape
var pGvRIJZpqdN
for (i = 0; i < 18000; i++)
    pGvRIJZpqdN = pGvRIJZpqdN + 0x77;
var s = "XY104CXY106FXY1072XY1065XY106DXY1020XY" + "1069XY1070XY1073\x75XY106DXY1020XY1064" + "XY106FXY106CXY106FXY1072XY1020XY1073XY" + "1069XY1074XY1020XY1061XY106DXY1065XY10" + "74\x25XY1020XY1063XY106FXY106EXY1073XY" + "1065XY1063XY1074XY1065XY1074\x75XY1072" + "XY1020XY1061XY1064XY1069XY10…";
pGvRIJZpqdN = unes(urpl(s));
\end{minted}
    \caption{Artificial malware example: obfuscated binary payload}
    \label{lst:obf3}
\end{listing}

The most common way for anti-virus software to identify PDF malware is to
search files for signatures or patterns of known malware. While cheap and
fast, signature-based methods are easily evaded through simple obfuscations. 

Indeed, all examples of PDF malware we examined obfuscate their JavaScript code
to avoid detection by matching the code's textual representation against a signature (or pattern).
While the vulnerable APIs that malware seeks to exploit might be well known, detecting them syntactically
can ultimately be prevented through obfuscation.
Code obfuscation in JavaScript can be achieved easily using the language's
built-in support for string manipulation, reflection, and different character-encoding.
\Cref{lst:obf1,lst:obf2,lst:obf3} give examples of PDF malware using different methods of obfuscation.
The first two examples use string concatenation and string encoding to cloak
the code they would execute and property name they seek to access. Both of these
obfuscations are simple enough to undo. The example in ~\Cref{lst:obf3} is more
complicated. It aliases and composes function objects, uses string
replacement, and the built-in \inlineJS{unescape} function to decode its malicious
payload. Through these powerful means of obfuscation, exploit code can be rewritten
(manually or automatically) in countless ways, and its detection requires the
signature database of anti-virus software to be updated continuously and promptly.
However, to reason about such complex code, more advanced techniques are needed.

To overcome the limitations of signature-based techniques, metadata~\cite{Smutz2012} 
and structure-based~\cite{Maiorca2012, Vsrndic2016} learning approaches have 
been proposed. Both types of approaches mainly differ in feature extraction.
While metadata-based approaches distinguish between benign and malicious
documents based on features such as file size, number of JavaScript components,
or number of embedded fonts, structure-based approaches classify documents 
based on paths in the PDF document tree. Because simple obfuscation techniques do not
alter the metadata or structure of PDF documents, such approaches proved to 
be very efficient at detecting PDF malware. However, because metadata and 
structural features do not cause malicious behavior, metadata and structure-based
approaches learn only what features are \emph{correlated} to malicious behavior. 
For this reason, they can be easily evaded through mimicry~\cite{Srndic2014} and 
reverse mimicry attacks~\cite{Maiorca2013} that hide malicious payloads in
files exhibiting metadata and structural properties of benign files.  

\begin{listing}
\begin{minted}{JavaScript}
var curDate = new Date();
var day = util.printd("ddd dd", curDate);
if (app.viewerVersion < 8.0) {
	/* trigger exploit */
}
\end{minted}
    \caption{Artificial malware example: exploit code active only on PDF viewers with a version less than 8.0}
    \label{lst:snd1}
\end{listing}

JavaScript-based detection approaches search for signs of malware
closer to the source of the problem, by targeting the JavaScript code embedded in PDF documents.
JavaScript-based detection approaches vary from fully static machine learning
approaches~\cite{Laskov2011,Vatamanu2012} to hybrid static and dynamic techniques~\cite{liu2014,Tzermias2011,Lu2013}. 
On the one hand, all the JavaScript-based machine learning approaches we are
aware of perform lexical analysis of the JavaScript code and thus capture only
lexical features of malicious PDFs, making them susceptible to mimicry 
attacks. Dynamic approaches (including the dynamic component of a hybrid approach), on the other hand, partly rely on either an
instrumented PDF viewer or a special JavaScript runtime environment to
dynamically detect malicious behavior. Any dynamic approach, however, is
inherently limited by the fact that they target a specific runtime
environment only.
They may miss malicious behavior due to simple checks that probe 
the environment, such as the one shown in \Cref{lst:snd1}.
As PDF viewers evolve and as vulnerabilities get fixed,
dynamic (and hybrid) approaches quickly become outdated unless significant effort
is invested in maintaining the analysis runtime environment.

\subsection{Our approach: \ourtool}
We propose the use of \emph{conservative} abstract interpretation of JavaScript
as a static analysis to detect malware in PDF documents.
By means of abstract interpretation, \ourtool hits a sweet spot in
the analysis landscape where it can \emph{statically}
consider all possible executions of the JavaScript code, and detect malicious
behavior without relying on a special runtime environment or requiring any user 
interaction. For example, during abstract interpretation, all event listeners (\eg
\mintinline{JavaScript}{Keystroke},
\mintinline{JavaScript}{Mouse Down},
\mintinline{JavaScript}{Mouse Enter})
are automatically triggered and analyzed.
Because it reasons about the runtime behavior of the JavaScript 
code instead of its structure or syntax, abstract interpretation is 
oblivious to mimicry attacks. On the other hand, because it is not tied to a 
specific runtime environment, \ourtool requires less maintenance 
than its dynamic analysis counterparts, and is not subject to sandbox evasion 
attacks. 

\begin{figure}[]
\centering
\colorlet{lcfree}{green!80!black}
\colorlet{lcnorm}{blue!80!black}
\begin{tikzpicture}[%
    >=triangle 60,              %
    start chain=going below,    %
    node distance=10mm and 40mm, %
    every join/.style={norm},   %
    ]
\tikzset{
  base/.style={draw, on chain, on grid, align=center, minimum height=4ex, font={\small}},
  proc/.style={base, rectangle, text width=7em},
  test/.style={base, diamond, aspect=2, text width=5em},
  term/.style={proc, rounded corners},
  coord/.style={coordinate, on grid, node distance=6mm and 25mm},
  nmark/.style={draw, cyan, circle, font={\sffamily\bfseries}},
  norm/.style={->, draw, lcnorm},
  free/.style={->, draw, lcfree},
  cong/.style={->, draw, lccong},
  it/.style={font={\small\itshape}}
}
\node [proc, join] (p1) {Extract JS code};
\node [test, join] (t101) {Success?};
\node [proc] (p01) {Abstract\\interpretation};
\node [test, join] (t1) {Fixpoint reached?};
\node [proc] (p2) {Semantic whitelist};
\node [test, join] (t2) {Benign\\{behavior}?};
\node [coord, right=of t1] (c1)  {}; \cmark{1}   
\node [coord, right=of t2] (c0)  {}; \cmark{0}   
\node [coord, left=of t2] (c2)  {}; \cmark{2}   

\node [coord, left=of p01] (c3) {}; \cmark{3}
\node [coord, right=of p1] (c5) {}; \cmark{5}
\node [coord, right=of t101] (c6) {}; \cmark{6}

\node [term, above=7mm of c5] (p0) {Input: PDF Document};
\node [coord] (c7) at (c5.north -| c3)  {}; \cmark{7};
\node [term, above=7mm of c7] (m0) {Input: \mbox{PDF-JS} Model};

\node [term, below=16mm of c2] (good) {Benign PDF};
\node [term, below=16mm of c0] (bad) {Malicious PDF};
\path (t101.south) to node [near start, xshift=1em] {$y$} (p01);
  \draw [*->,lcnorm] (t101.south) -- (p01);

\path (t1.south) to node [near start, xshift=1em] {$y$} (p2);
  \draw [*->,lcnorm] (t1.south) -- (p2);

\draw [->,lcnorm] (p0.south) -- (c5) -- (p1.east);
\draw [->,lcnorm] (m0.south) -- (c3) -- (p01);

\path (t1.east) to node [near start, yshift=1em] {$n$} (c1); 
  \draw [o->,lcfree] (t1.east) -- (c1) -- (bad);
\path (t2.east) to node [near start, yshift=1em] {$n$} (c0); 
  \draw [o->,lcfree] (t2.east) -- (c0) -- (bad);
\path (t2.west) to node [near start, yshift=1em] {$y$} (c2); 
  \draw [o->,lcfree] (t2.west) -- (c2) -- (good);
\path (t101.east) to node [near start, yshift=1em] {$n$} (c6); 
  \draw [o->,lcfree] (t101.east) -- (c6) -- (bad);
\end{tikzpicture}

 \caption{Overview of our malware detection analysis}
\label{fig:main}
\end{figure}
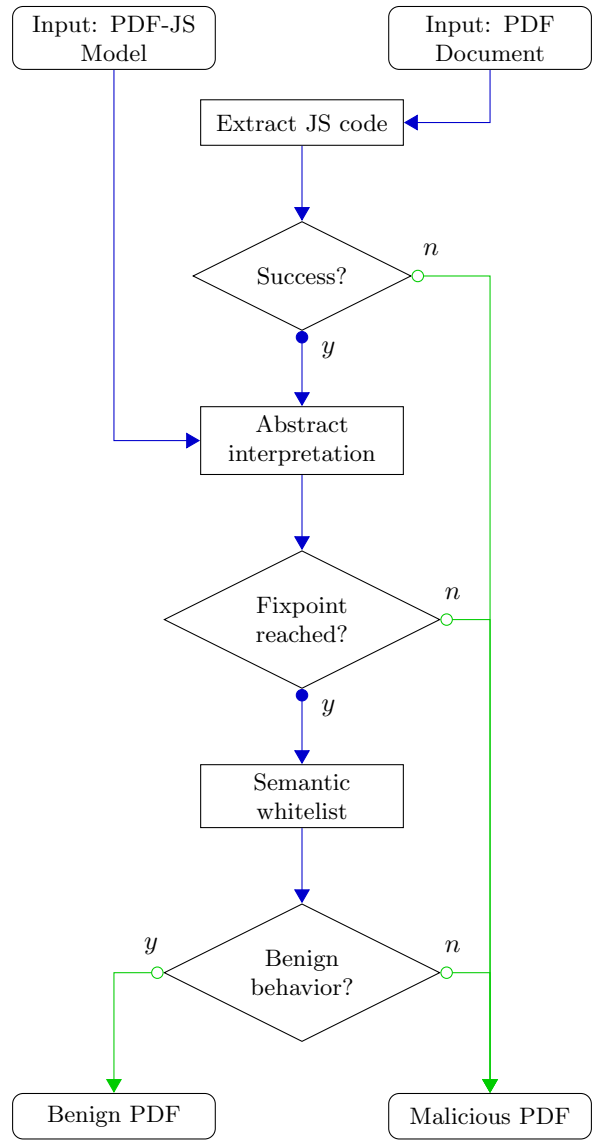

By conservative, we mean that when in doubt, our analysis will err on the safe
side. In other words, we are willing to accept that our analysis may regard a
harmless PDF as malicious (i.e., a \emph{false positive}), but we do not accept
the opposite (i.e., a \emph{false negative}).
Abstract interpretation considers all possible behavior of the code under
analysis without directly executing it.
Based on its result, we use a whitelisting mechanism to allow only safe
behavior (\eg restricting the use of JavaScript APIs to a known safe subset)
and reject everything else as malware.

The flowchart in \Cref{fig:main} highlights
the main steps of \ourtool, our PDF malware detection tool.
\ourtool starts by extracting the JavaScript code from the input document.
Then, it complements the extracted code with a model of the
JavaScript runtime environment inside a PDF viewer following Adobe's specification, hereafter called the \emph{PDF-JS} model, and performs 
abstract interpretation.
If abstract interpretation does not complete (i.e., it cannot reach a
fixpoint) within a certain time, the document is immediately reported as
malicious.
If abstract interpretation terminates, \ourtool then
checks if the document exhibits potentially malicious behavior. If so, the
document is reported as malicious. Otherwise, it is reported as benign.

\subsection{Background: Abstract Interpretation}
Abstract interpretation is a mathematically well-founded framework for static
analysis introduced by Cousot and Cousot in~\cite{Cousot1977}. It addresses the
challenge of computing non-trivial properties of a program, which is known 
to be undecidable when the concrete language semantics is used (c.f., Rice's 
theorem~\cite[chapter~9]{Hopcroft2001}). When concrete values and operations are 
approximated with abstract values and abstract operations, however,
such an abstract interpretation of a program becomes computable.
This comes at the cost of losing precision for some properties of a program due
to the abstraction (approximation) that is applied.
In the context of static analysis, such a (partially) imprecise result means that
some of our questions about a program (e.g., \emph{``is it malicious?}'') have to be
answered with \emph{``We don't know''} for the analysis to be sound.
In the following paragraphs we explain some of the concepts behind abstract
interpretation that are required later on, and we give a step-by-step example.
A formal introduction to abstract interpretation can be found
in~\cite[chapter~4]{NNH99}.

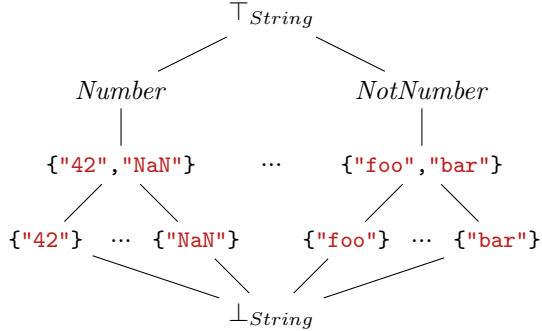
\begin{figure}
\centering
\begin{tikzpicture}
\node(top) at (0,4) {$\top_{String}$};
\node(number) at (-2,3) {\textit{Number}};
\node(notnumber) at (2,3) {\textit{NotNumber}};
\node(42nan) at (-2,2) {\mintinline{JavaScript}{{"42","NaN"}}};
\node(genSet) at (0,2) {...};
\node(foobar) at (2,2) {\mintinline{JavaScript}{{"foo","bar"}}};
\node(42) at (-3,1) {\mintinline{JavaScript}{{"42"}}};
\node(genNumber) at (-2,1) {...};
\node(nan) at (-1,1) {\mintinline{JavaScript}{{"NaN"}}};
\node(foo) at (1,1) {\mintinline{JavaScript}{{"foo"}}};
\node(genString) at (2,1) {...};
\node(bar) at (3,1) {\mintinline{JavaScript}{{"bar"}}};
\node(bot) at (0,0) {$\bot_{String}$};
\draw(bot) -- (42);
\draw(bot) -- (nan);
\draw(bot) -- (foo);
\draw(bot) -- (bar);
\draw(42) -- (42nan);
\draw(nan) -- (42nan);
\draw(42nan) -- (number);
\draw(foobar) -- (notnumber);
\draw(foo) -- (foobar);
\draw(bar) -- (foobar);
\draw(number) -- (top);
\draw(notnumber) -- (top);
\end{tikzpicture}
\caption{SAFE's string abstract domain. It keeps track of sets of at most $k$  different (number-)~strings
before approximating them as \textit{Number} or \textit{NotNumber}. ($k=2$ in this figure.)}
\label{fig:safeLattice}
\end{figure}

From a program analysis point of view, abstract interpretation gives us the ability
to statically (\ie without running the program) build an abstract state for
every point in the program. These states capture information about possible
concrete executions and we can use them to
validate assertions or find problems in the programs we want to analyze.
It is up to us, as the designer of the abstract interpretation analysis, to choose an
appropriate abstraction.
An abstraction consists of (1) abstract domains and (2) abstract semantics. The
abstract domains capture the program states in our analysis as abstract values.
It is important to note that an abstract value must be able to represent more
than one concrete value at a time.
A domain maintains information in the form of a \emph{lattice}, which may have
up to infinite elements and infinite height.
(See \Cref{fig:safeLattice} for an example of the abstract
domain for string values used by \ourtool; this is a simple \emph{powerset lattice}
ordered by \emph{set inclusion}.)
Depending on the abstraction function, a domain can capture information about possible concrete values using wide
approximations. For example, if we were interested only to find out whether a
numeric value can be negative, we could abstract an integer value as its
sign; or we abstract it as a set or a range of concrete integer values.
We also need to encode the semantics of abstract operations, which approximates
concrete operations on abstract values.
Given an abstraction, we compute a fixpoint for a program in the abstraction.
Note that in practice, analysis might not reach a fixpoint for all programs 
within feasible time and memory bounds.

\begin{listing}
\begin{minted}{JavaScript}
var msg = "hello world"; |$\label{ai:hello}$|
// local: msg := "hello world"
var arr = [ |$\label{ai:arr}$|
  function(s) { console.println(s); },
  function(s) { doc.getField("msgField").value = s; }
];
// heap: #1 := Func
//       #2 := Func
//       #3 := Obj { "0" := #1, "1" := #2, length := 2 }
// local: arr := #3, msg
var i = doc.getField("inputField");|$\label{ai:getField}$|
// local: arr, i := $\top_{String}$, msg
var n = Number.parseInt(i);|$\label{ai:parseInt}$|
// local: arr, $i$, msg, n := $\top_{Number}$
var fn = arr[n];|$\label{ai:lookup}$|
// local: fn := {#1, #2, undefined}
if (fn === undefined)|$\label{ai:if}$|
  // local: fn := {undefined}
  app.alert("Unexpected input: " + i);|$\label{ai:alert}$|
  // calls app.alert()
else
  // local: fn := {#1, #2}
  fn(msg); |$\label{ai:else}$|
  // calls either #1 or #2
\end{minted}
        \caption{PDF JavaScript snippet with abstract interpretation state in comments}
    \label{lst:ai-example}
\end{listing}

\newcommand{\aiAbs}[1]{\emph{#1}}
\newcommand{\lref}[1]{l.~\ref{ai:#1}}
\minorsubsection{Abstract interpretation example}
We present a simple JavaScript code snippet in \Cref{lst:ai-example}. 
Lines that start with \inlineJS{//} show updates to the
abstract state after every instruction.
The \inlineJS{var} keyword at line \ref{ai:hello} creates a local
variable, which our abstract state stores directly,
together with the right-hand side string-value, in the current \aiAbs{local} scope.
The array \inlineJS{arr} (created at line \ref{ai:arr}) contains two function objects, which are
stored in the abstract heap at address \aiAbs{\#1} and \aiAbs{\#2} respectively.
The abstract array object itself is stored at address \aiAbs{\#3} and referenced by a
new variable in \aiAbs{local}. The \inlineJS{arr} object contains three properties: \aiAbs{``0''} and
\aiAbs{``1''} point to the function objects at their respective addresses; the
internal \aiAbs{length} property approximates the number of items contained in
an array (or any JavaScript object).
The next two instructions (lines \ref{ai:getField} and \ref{ai:parseInt}) call
a PDF API and JavaScript built-in function respectively.
The call to \inlineJS{getField} returns a user input that cannot be known
statically. But based on the specification of \inlineJS{getField}, our abstract
interpretation can approximate the returned value with the abstract value
$\top_{String}$, which represents \emph{any string}. This value is then passed to
\inlineJS{parseInt}~(\lref{parseInt}), for which our analysis has a semantic model:
it returns an integral number if the string argument can be parsed and
\inlineJS{NaN} otherwise.
In this case, for \emph{any string} as input, it returns $\top_{Number}$, \ie
\emph{any number}.
The resulting value stored in \inlineJS{n} is then used in a property lookup on the array object
\inlineJS{arr}~(\lref{lookup}). The information in our abstract state at this
point is precise enough to give a close approximation for the lookup with a key
value of $\top_{Number}$, which is converted to a string value during lookup according
to JavaScript semantics. Consequently, the values with property names \aiAbs{``0''} and
\aiAbs{``1''} match the lookup because they are number strings, and are thus
returned as results to \inlineJS{fn}. The value \inlineJS{undefined} is also returned
because $\top_{Number}$ includes numbers that are not in \aiAbs{arr} (\eg $2, 42.3, ...$).
Note, however, that because $\top_{Number}$ matches number strings only, the lookup does 
not match \aiAbs{length} or any of the internal array functions (e.g. \inlineJS{find()}), 
which can be accessed from a JavaScript array object by following its prototype chain.
To approximate the control-flow behavior of the if-statement starting in line
\ref{ai:if}, we perform its test against our abstract state.  Abstract
interpretation determines that because \inlineJS{fn} may be \inlineJS{undefined}, the
then-branch~(\lref{alert}) can be taken, and the call to \inlineJS{app.alert}
may be executed.  In the else-branch (also reachable due to the abstract value of \inlineJS{fn}),
abstract interpretation can remove \inlineJS{undefined} from the possible values of \inlineJS{fn} (before reaching line~\ref{ai:else})
because it contradicts the if-condition.
This proves that the call to the function object \inlineJS{fn}, can invoke only
the two functions defined inside \inlineJS{arr}. It cannot call any other
function, and cannot fail due to \inlineJS{fn} not being a function object.

\section{Malware Detection}
\label{sec:ai}
In this section we describe in detail how to detect malicious JavaScript in PDF files
using conservative abstract interpretation and how we overcome parser confusion attacks 
that plague existing PDF JavaScript code extractors. Our description of processing steps 
follows the flowchart in \Cref{fig:main}.
\subsection{Pre-processing step: extraction}
\label{sec:extraction}

Our approach requires JavaScript code to be extracted from PDF documents before it can be analyzed.
While conceptually simple, the PDF format makes extraction extremely tricky. Indeed, JavaScript
code can be embedded in different PDF constructs, encoded with various uncommon encodings, compressed, and 
encrypted, meaning that JavaScript code extraction requires a full-fledged PDF parser. Moreover, Carmony 
et al.~\cite{carmony2016} showed that PDF viewers often deviate from the specification, in an attempt to
``just work'', and that existing open-source and commercial JavaScript code extractors all fail
to extract code from various PDF documents. As a result of their work, they compiled a set of 2952 
PDF documents that are known to cause extraction issues in one or more JavaScript code extractors. Starting 
from those documents with hard-to-extract JavaScript code, we extended an
existing commercial extractor~
\cite{WebCleanContent}
until it could successfully extract JavaScript code from \emph{all} documents in the set. Because static code extraction
can reach code that might not be loaded dynamically (e.g form actions that are only triggered when
interacting with the document), we claim that our approach analyzes a strict superset of the code that 
can be extracted by dynamically loading a PDF document in a sandbox.

\begin{table}
 \centering
 \caption{Extractor limitations and associated PDF constructs}
 \begin{tabular}{ll}
   \toprule
   Extractor limitation & Problematic PDF construct \\
   \midrule
   \multirow{10}[6]{*}{Implementation Bugs} & Comment in document trailer \\
   \cmidrule{2-2}
   & Comment in dictionary object \\
   \cmidrule{2-2}
   & Trailing whitespace in stream data \\
   \cmidrule{2-2}
   & \textbf{Null object reference} \\
   \cmidrule{2-2}
   & Security handler revision 5 hex \\ 
   & encoded encryption data parsing \\
   \cmidrule{2-2}
   & Security handler revision 3, 4 \\
   & encryption key computation \\
   \cmidrule{2-2}
   & Hexadecimal string literal in encoded \\
   & objects \\
   \midrule
   \multirow{3}{*}{Design Errors} & Use of orphaned encryption objects \\
   \cmidrule{2-2}
   & Security handler revision 5 key \\
   & computation with clear metadata \\
   \midrule
   \multirow{3}{*}{Omissions} & No XFA support \\
   \cmidrule{2-2}
   & No security handler revision 5 support \\
   \cmidrule{2-2}
   & No security handler revision 6 support \\
   \midrule
   \multirow{6}{*}{Ambiguities} & Invalid object keywords \\ 
   \cmidrule{2-2}
   & No cross-reference table \\
   \cmidrule{2-2}
   & \textbf{Wrong or missing entries in the} \\
   & \textbf{cross-reference table} \\
   \cmidrule{2-2}
   & \textbf{Partially broken compressed} \\
   & \textbf{streams} \\
   \bottomrule
 \end{tabular}
 \label{tab:parser_errors}
\end{table}

In the following paragraph, we briefly introduce the PDF format, focusing only on elements that are 
relevant to code extraction. We refer the interested reader to~\cite{GuillaumeEndignoux2016} for an
extensive description of the Portable Document Format (PDF). For code extraction
purposes, the four most important elements of the PDF syntax are: (1) direct objects, which are the basic
building blocks of a PDF; (2) indirect objects, which are uniquely identified, and can be referenced from 
elsewhere in the document; (3) cross-reference tables, which contain the positions of objects in the file;
and (4) content streams, which store various parts of the document content. Content streams are composed
of two parts: a stream that is an optionally compressed and encrypted byte sequence, and a meta-data dictionary object that 
carries information about the stream's encoding and how to uncompress it.  Because Adobe Reader can cope with
partially broken compressed streams, and unspecified encodings, we extended our extractor with the
same capabilities.

Furthermore, because JavaScript code in PDF documents is usually broken into snippets and spread across several 
content streams, a code extractor must not only extract the various snippets from streams, it must also parse 
the document in order to recover its structure and re-assemble the snippets into a semantically valid program.
In~\cite{carmony2016}, the authors list several constructs that are known to cause PDF parser failures and
extraction errors. We report those constructs in Table~\ref{tab:parser_errors}, along with additional 
problematic constructs we addressed in our extractor (in bold). In Table~\ref{tab:parser_errors}, security
handlers refer to various encryption algorithms that can be used to encrypt streams, and XFA refers
to the XML Forms Architecture, which is supported by the PDF specification, and that allows the embedding of 
JavaScript actions in XML forms. After meticulous extensions, our extractor now supports all constructs listed 
in Table~\ref{tab:parser_errors} and extracts JavaScript code from \emph{all} of the original 2952 PDF 
documents with hard-to-extract JavaScript code that contain non-empty JavaScript code and that do not
cause our extractor to fail. Indeed, during manual investigation of the documents with no extracted JavaScript, we observed that all of them contain an 
empty JavaScript string (e.g. \texttt{/JS()/S/JavaScript}). We believe that because the approach in~\cite{carmony2016} 
detects JavaScript code by monitoring loads of the \texttt{EScript.api} module, which would be triggered by the 
\texttt{/JavaScript} keyword, it mistakenly tags those files as containing JavaScript code. We also observed
that most of the hard-to-extract files that cause extraction failure are so broken that they cannot be opened with
Acrobat Reader DC 2018.011. Hence, we are highly confident that our extractor retrieves all the JavaScript code 
that can be extracted.

\subsection{Main analysis step: abstract interpretation}
During the main analysis step, we perform abstract interpretation of the extracted
JavaScript code. 
As secondary input, we provide a model of the JavaScript
runtime environment emulating that of a concrete PDF viewer, which
we refer to as our \emph{PDF-JS model}.
The role of the PDF-JS model is to provide extracted JavaScript code with an
abstract environment for analysis emulating that of a PDF viewer application.
It thus captures (a subset of) the PDF-JavaScript specification~\cite{AdobePDFAPI,AdobePDF3DAPI}.
For example, the global static objects \mintinline{JavaScript}{app} and
\mintinline{JavaScript}{doc} are made available as part of the JavaScript
environment according to Adobe's documentation.
Unlike a concrete JavaScript-based environment, which would be used for dynamic analysis,
our model can make use of
abstract semantics (i.e. not all API functions must provide concrete results),
as long as it remains conservative, i.e., it must not under-approximate the
behavior of the JavaScript API.
We present the PDF-JS model in detail in \Cref{sec:model}.

To support JavaScript code in XFA and its interactions with objects specified
as XML entities, we provide additional modeling.
In accordance with available documentation~\cite{xfa}, the analysis extracts
and dynamically models XFA entities as JavaScript objects and, using the same
principles as the PDF-JS model, provides an environment to analyze XFA
JavaScript code.

The result we receive from this analysis step must represent an over-approximation of
the JavaScript code's behavior.
In situations where abstract interpretation cannot reach a fixpoint,
no valid over-approximation is available, and thus the analysis immediately reports a potential malware.
Causes for not reaching a fixpoint are:
(1) the analysis reaching a timeout or exhausting the available memory;
and (2) syntactic or semantic errors in the extracted JavaScript code causing the analysis to fail.
Only if the analysis reaches a fixpoint, we pass on the result of abstract interpretation to
the final \emph{whitelisting} step.
\subsection{Post-processing step: semantic whitelist}
The last step of our static analysis classifies the extracted JavaScript as either \emph{safe} or \emph{malicious}.
This is done by inspecting the result of abstract interpretation from the previous step.
For our analysis to be conservative, \ourtool has to reject a PDF document as
malicious if the result of performing abstract interpretation on its extracted
JavaScript code cannot prove the absence of all of the following:
\begin{enumerate}
  \item a call to a vulnerable API method \label{it:API}
  \item a potentially malicious program behavior \label{it:behavior}
  \item an \emph{unknown behavior} \label{it:unknown}
\end{enumerate}
We detect the use of vulnerable APIs (\labelcref{it:API}) by building a PDF-JS model (see \Cref{sec:model}), which,
through semantic modeling, selectively whitelists those API methods known \emph{not} to be vulnerable.
As a result, any call to a non-whitelisted method is detected as malicious.
To detect the second class of malware (\labelcref{it:behavior}) that typically performs heap spraying
to exploit memory corruption in the language runtime, we detect the 
creation of large values. Specifically, our semantic whitelisting detects the following potentially 
malicious program behavior:
\begin{itemize}
\item string length exceeding a predefined limit;
\item object (array) size exceeding a predefined limit.
\end{itemize}
Finally (\labelcref{it:unknown}), because \ourtool is
conservative, it needs to report any code as malware that exhibits \emph{unknown
behavior}, which makes it impossible to prove the absence of calls to vulnerable API methods
or malicious program behavior.
For example, calls to all 
\mintinline{JavaScript}{eval}-like functions that allow arbitrary strings to be 
interpreted as code are causes of unknown behavior. Indeed, the effect of calling
an \mintinline{JavaScript}{eval}-like function is usually statically intractable.
Similarly, calls to imprecise function objects (due to aliasing or function
lookup using an unknown input value, \eg a $\top_{String}$ value)
cause unknown behavior as well.

\bigskip
\subsection{The PDF-JS model}
\label{sec:model}
Our analysis depends on a model that emulates the PDF environment during
abstract interpretation.
In our implementation, this model is based on a set of PDF documents containing benign JavaScript
code, while relying on an API reference~\cite{AdobePDF3DAPI, AdobePDFAPI} and
being aware of reported vulnerabilities.
Because \ourtool is conservative, it always interprets calls to non-whitelisted 
functions as malicious. As a result, given valid JavaScript code, reducing the 
false positive rate of \ourtool usually means extending the model. This is a 
straight-forward and incremental process, and we show in \Cref{sec:ExpSetup} that it 
yields very good results in practice.

\begin{listing}[t]
\begin{minted}{JavaScript}
function PDF_Event() {
   this.type = |$\top_{String}$|
   this.name = |$\top_{String}$|
   this.change = |$\top_{String}$|
   this.changeEx = |$\top_{String}$|
   this.commitKey = |$\top_{Number}$|
   this.fieldFull = |$\top_{Bool}$|
   this.keyDown = |$\top_{Bool}$|
   this.modifier = |$\top_{Bool}$|
   this.rc = |$\top_{Bool}$|
   this.selEnd = |$\top_{Number}$|
   this.selStart = |$\top_{Number}$|
   this.shift = |$\top_{Bool}$|
   this.source = PDF_DOM_NODE
   this.target = PDF_DOM_NODE
   this.targetName = PDF_DOM_NODE.name
   this.value = |$\top_{String}$|
   this.willCommit = |$\top_{Bool}$|
}

var PDF_DOM_NODE = {
    name: |$\top_{String}$|
    setFocus: function() { return |$\top_{Bool}$| },
    value: |$\top_{String}$|
}
\end{minted}
    \caption{Mock PDF event object used for analysis by \ourtool}
    \label{lst:pdfEvent}
\end{listing}

\emph{Unknown} inputs
can lead to \emph{unknown behavior} during abstract interpretation, depending on how they are used.
This can result in a
non-malicious PDF document being classified as malicious (\ie a \emph{false positive}).
To reduce false positives, we can optionally enrich the PDF-JS model with
concrete metadata extracted from the PDF document.
Inputs from the user and the host environment, however, cannot
be statically known and thus always introduce \emph{unknown} values during abstract interpretation.
Depending on their origin, however, such non-deterministic values can be modeled with different levels of precision.
For example, an input from the user in a free text form field would be conservatively modeled
as $\top_{String}$ while the runtime variable representing the operating system of the user can be
modeled as a set of concrete values like \mintinline{JavaScript}{{'WIN','MAC','UNIX'}}.
In general, we can start with a model containing many conservative approximations and refine it where necessary.

Similar to JavaScript in web browsers, JavaScript code in PDF documents is largely event-driven,
\ie either triggered by system or user events. While we observed that most malicious code is executed when the PDF 
document is opened, JavaScript code may be placed in event handlers, where it is executed only
on certain user actions (\eg clicking a form field). Because we cannot exclude the possibility that
an event handler might contain malicious code, \ourtool must consider every
event handler as an entry point. Moreover, because event handlers can have side effects that alter
the computation of other event handlers, \ourtool loops over all handlers and triggers 
them until a fixpoint is reached, indicating that it computed a suitable over-approximation for the
whole program.
Because event handlers receive an \mintinline{JavaScript}{event} object as argument and operate
on its properties, our PDF-JS model defines a mock PDF event object that is passed to event
handlers at analysis time. \Cref{lst:pdfEvent} shows the mock PDF event object of our PDF-JS
model. The right-hand side values $\top_{String}$, $\top_{Bool}$, $\top_{Number}$ represent abstract values.
They stand for \emph{any} string, \emph{any} Boolean, and \emph{any} number respectively.
The \mintinline{javascript}{PDF_DOM_NODE} abstracts nodes in the PDF Document Object Model (DOM) to which
events can be attached.

\section{Experimental Evaluation}
\label{sec:ExpSetup}

To assess the effectiveness, and robustness of our technique, we
implemented the \ourtool tool, and investigated
the following research questions:

\noindent\textbf{RQ1:} How does the precision, recall and accuracy of 
\ourtool compare to state-of-the-art malware detection tools?

\noindent\textbf{RQ2:} How resilient is \ourtool to parser
confusion and mimicry attacks compared to state-of-the-art tools?

\subsection{Experiment setup}
\label{sec:ExpSetupSub}

\ourtool is based on version $1.0$ of the
\toolname{SAFE} abstract interpretation framework for ECMAScript~\cite{lee2012}.
We complement the original \toolname{SAFE} framework with
(1) malware-specific analyses,
(2) our semantic models for the PDF JavaScript and XFA environments,
and (3) further modeling of the interactions between XML and JavaScript in XFA forms.

Experiments were conducted on a set of $14\,306$ benign and $9410$ malicious 
PDF documents.
Malicious samples were collected from VirusShare~\cite{VirusShare},
a free online repository of malware samples, Contagio~\cite{Contagio},
and VirusTotal~\cite{VirusTotal}.
The benign benchmark set contains non-malicious PDF documents from Contagio,
VirusTotal, PDF attachments from a public email dataset~\cite{EnronEDRM}, as
well as samples collected from the Web (from Google
queries targeting PDF documents, e.g. \texttt{filetype:PDF}), test cases for
\toolname{PDF.js}, the PDF-rendering engine of Mozilla~\cite{pdfjs}, test cases
for \toolname{PDFium}, the PDF-rendering engine of Chrome~\cite{pdfium}, and
interactive documents from the pdfPictures~\cite{pdfpictures} website.
The samples from VirusTotal include the hard-to-extract set of
documents\footnote{\url{https://goo.gl/qtbuOC}} from \cite{carmony2016}, which
we refer to in \Cref{sec:extraction}.
All downloaded PDF documents were confirmed to be non-malicious using VirusTotal.
\Cref{tab:benchmark-stats}
lists the different benchmark sets, the number of files they contain, and
whether they contain malicious or benign samples.

\begin{table}[t]
 \centering
 \caption{PDF sample benchmarks used in this study}
  \begin{tabular}{lll}
    \toprule
    Benchmark & Source & \#Files\\
    \midrule
    \multirow{3}{*}{Malicious}%
    \begin{filecontents*}{data/sample-count-bad.csv}
Benchmark,Count
Contagio,7110
VirusShare,1206
VirusTotal,1094
\end{filecontents*}
\csvreader[late after line=\\]{data/sample-count-bad.csv}{}{&\csvcoli & \csvcolii}
    \midrule
    \multirow{4}{*}{Benign}%
    \begin{filecontents*}{data/sample-count-good.csv}
Benchmark,Count
Contagio,388
EDRMv2,4434
VirusTotal,2137
Web,7347
\end{filecontents*}
\csvreader[late after line=\\]{data/sample-count-good.csv}{}{&\csvcoli & \csvcolii}
    \bottomrule
  \end{tabular}
 \label{tab:benchmark-stats}
\end{table}

To extract JavaScript code from PDF documents, we extended
version $2015.1.4$ of the \toolname{Clean Content} SDK~\cite{WebCleanContent},
as described in \Cref{sec:extraction}.
In the rare cases where
a PDF document causes \CC to fail with an extraction error,
we pre-process the document with a modified version of \toolname{PDFBox}~\cite{pdfbox} in an
attempt to fix structural issues.
Also, we syntactically remove one particular nonsensical but benign code
fragment, a call%
\footnote{\inlineJS{jQuery.post(Drupal.settings.basePath + 'jstats.php', {...})}}
to the non-existent \inlineJS{jQuery.post} method, from the extracted
JavaScript code.
Instances of this call were introduced into our web-sourced dataset by an obviously broken web-based PDF creator.

For our experimental evaluation, we set a $30$ second timeout for the abstract
interpretation step, after which \ourtool rejects an input as malware.
We run six instances of \ourtool in parallel sharing eight cores of a Xeon
E5-2.60GHz with 32GB RAM.
On average \ourtool takes less than $4$ seconds to perform
analysis (\ie extraction and abstract interpretation) of a single PDF document.

The PDF-JS model in \ourtool was incrementally extended to
whitelist the subset of JavaScript functionality used by the benign samples in our benchmarks. 
Our experience suggests that supporting a new functionality in the PDF-JS model typically
requires one to two lines of JavaScript code. In cases where a fine-grained model is
needed, however, a developer can use all (sane) features of the JavaScript
language.

\subsection{RQ1: Comparison to state-of-the-art tools}

\begin{table}
\centering
\caption{Numbers of true positives (TP), true negatives (TN), false positives (FP), false negatives (FN), and errors per tool}
\begin{tabular}{llllll}
  \toprule
  Tool & TP & TN & FP & FN & Errors \\
  \midrule
  \begin{filecontents*}{data/raw_counts.csv}
Tools,TP,TN,FP,FN,errors
Slayer,9125,13877,10,72,631
Hidost,9215,14086,6,31,377
{\ourtool},9396,13920,321,7,72
\end{filecontents*}
\csvreader[late after line=\\]{data/raw_counts.csv}{}{\csvcoli & \csvcolii & \csvcoliii & \csvcoliv & \csvcolv & \csvcolvi}
\bottomrule
\end{tabular}
\label{tab:comparison}
\end{table}

\begin{table}
\centering
 \caption{Precision, recall and accuracy of \toolname{Slayer}, 
   \toolname{Hidost}, and \ourtool}
 \begin{tabular}{lccc}
   \toprule
   Tool & Precision & Recall & Accuracy\\
   \midrule
   \begin{filecontents*}{data/pre_rec_acc.csv}
Tools,Precision,Recall,Accuracy
Slayer,95.62,99.23,97.89
Hidost,97.72,99.67,98.95
{\ourtool},96.06,99.93,98.34
\end{filecontents*}
\csvreader[late after line=\\]{data/pre_rec_acc.csv}{}{\csvcoli & \csvcolii & \csvcoliii & \csvcoliv}
   \bottomrule
\end{tabular}
\label{tab:metrics}
\end{table}

To determine how \ourtool compares to the state-of-the-art, we compared the detection rate of \ourtool against two other publicly available PDF
malware detection tools: \toolname{PDF Malware Slayer}~\cite{Maiorca2012}, and \toolname{Hidost}~\cite{Vsrndic2016}.

\toolname{PDF Malware Slayer} first identifies keywords that are characteristic of benign 
and malicious documents from sets of benign and malicious PDFs. It then trains a \toolname{Random Forests}
classifier on feature vectors obtained by computing the frequency of characteristic keywords 
in each document. To measure the precision, recall, and accuracy of \toolname{PDF Malware Slayer},
we perform a 10-fold cross-validation experiment with default parameters and report averaged results. 

\toolname{Hidost} also uses a \toolname{Random Forests} classifier to identify malicious PDFs.
\toolname{Hidost} mainly differs from \toolname{PDF Malware Slayer} in the way it extracts feature
vectors from PDFs. \toolname{Hidost} builds feature vectors by extracting structural paths from PDF 
documents, where structural paths capture the embedding of PDF components. 
Because \toolname{Hidost} was initially trained and tested on a very large dataset, comprising more 
than 400,000 documents, it considers only structural paths present in at least 1000 documents by 
default. To accommodate our smaller dataset, we reduced this threshold to 200. Because \toolname{Hidost} 
also uses a random classifier, we perform a 10-fold cross-validation experiment and report averaged 
results.

\Cref{tab:comparison} summarizes the output of \toolname{PDF Malware Slayer},
\toolname{Hidost} and \ourtool on our benchmarks. In \Cref{tab:comparison},
the last column counts ``errors'', \ie when the tool failed to analyze a PDF
document, either by failing silently, ignoring it, or by exiting with an error.
The errors in \ourtool all stem from heavily broken PDF documents that 
cause our code extractor to fail.

\Cref{tab:metrics} lists the precision, recall, and accuracy of
all of the investigated tools. Equations (\ref{eqn:precision}) to (\ref{eqn:acc})
show the corresponding formulas, where $TP$, $FP$, $TN$ and $FN$ stand for
true positive, false positive, true negative and false negative respectively.

\begin{align}
Precision = \frac{TP}{(TP + FP)}
\label{eqn:precision}\\
Recall = \frac{TP}{(TP + FN)}\\
Accuracy = \frac{(TP + TN)}{(TP + TN + FP + FN)}
\label{eqn:acc}
\end{align}

All of the evaluated tools failed to analyze some PDF documents, either through silent failure 
(\eg not analyzing the document) or by exiting with an error. Because \ourtool 
is conservative, we treat any extraction failure as an indication that the document
is malicious. To perform a fair comparison, we treat errors in other tools as malicious reports too.

Going back to our initial research question,
the metrics presented in \Cref{tab:metrics} show that \ourtool achieves comparable 
precision, recall and accuracy with state-of-the-art PDF malware detectors.

\subsection{RQ2: Resilience to evasion attacks}

\begin{table*}
  \caption{Parser confusion and reverse mimicry attacks}
  \begin{tabular}{p{7cm}ccc}
   \toprule
     \multirow{2}{*}{Obfuscation} &  \multirow{2}{*}{Slayer} & \multirow{2}{*}{Hidost} & \multicolumn{1}{p{1cm}}{\centering SAFE-\\PDF}\\
   \midrule
   None & \cmark & \cmark & \cmark \\
   \hangindent=1em Flate compression, obj streams & \cmark & \cmark & \cmark \\
   \hangindent=1em Flate compression, R5 security handler & \xmark & \cmark & \cmark \\
   \hangindent=1em Flate compression, R5 security handler, obj streams & \xmark & \cmark & \cmark \\
   \hangindent=1em Flate compression, R6 security handler & \xmark & \xmark & \cmark \\ 
   \hangindent=1em Flate compression, R6 security handler, obj streams & \xmark & \xmark & \cmark \\
   \hangindent=1em Flate compression, R6 security handler, obj streams, comment in trailer & \xmark & \xmark & \cmark \\
   \hangindent=1em JS encoded as UTF-16BE in hex string & \cmark & \cmark & \cmark \\
   \hangindent=1em JS encoded as UTF-16BE in hex string, flate compression, obj streams & \cmark & \cmark & \cmark \\
   \hangindent=1em JS encoded as UTF-16BE in hex string, flate compression, R5 security handler, obj streams, comment in trailer & \xmark & \cmark & \cmark \\
   \midrule
   Reverse mimicry attack + parser confusion & \xmark & \xmark & \cmark \\
   \bottomrule
\end{tabular}
\label{tab:confusion}
\end{table*}

To evaluate the resilience of \ourtool to evasion, we evaluated \ourtool on 
malicious PDF documents that were specifically designed by the 
authors of \cite{carmony2016} to evade detection by performing parser confusion and reverse 
mimicry attacks~\cite{Maiorca2013}. Table~\ref{tab:confusion} 
shows how many evasive variants were detected by \toolname{Slayer}, \toolname{Hidost}, and 
\ourtool. Interestingly, while parser confusion attacks were designed to primarily target 
approaches that rely on JavaScript code extraction, Table~\ref{tab:confusion} shows
how tools that rely on structural PDF features are also affected. Indeed, all malicious 
documents containing R5 handlers evaded \toolname{Slayer}, while documents containing R6
security handlers evaded both \toolname{Slayer} and \toolname{Hidost}. \ourtool
caught all evasive variants based on parser confusion.

Furthermore, because \ourtool statically analyzes program behavior, it is oblivious 
to reverse mimicry attacks, which are known to be very effective against structure-based 
approaches~\cite{Maiorca2013,Srndic2014,xu2016}. 
Indeed, as shown in Table~\ref{tab:confusion}, \ourtool could detect the malicious
payload, even in the presence of both reverse mimicry and parser confusion attacks.
\section{Discussion}
\label{sec:discussion}

In this section, we discuss open challenges for JavaScript-based PDF
malware detectors and present a threat model for \ourtool
together with possible attacks.

\subsection{JavaScript-based malware detection}

All JavaScript-based PDF malware detectors need to extract the JavaScript code 
from the PDF document either dynamically through an instrumented PDF viewer
or statically using a stand-alone extractor. As highlighted in~\cite{carmony2016},
however, because the PDF specification is very complex, extracting JavaScript 
code from PDF documents is far from trivial. While we addressed all
known limitations of existing extractors, static analysis of JavaScript remains
very difficult. In the following paragraphs, we discuss the challenges that
\ourtool faced during analysis of our benchmark set.

\begin{table}
  \centering
  \caption{Causes of malware reports by \ourtool}
  \begin{tabular}{lllS}
    \toprule
    Benchmark & Report Cause & Count & {Percentage}\\
    \midrule
    \multirow{5}{*}{Benign}%
    \begin{filecontents*}{data/causes-good.csv}
cause,count,percent
Unexpected behavior,145,37.56476683937824
Malicious behavior,129,33.41968911917099
Extraction error,65,16.83937823834197
JS parsing error,41,10.621761658031089
Fixpoint not reached,6,1.5544041450777202
\end{filecontents*}
\csvreader[late after line=\\]{data/causes-good.csv}{}{& \csvcoli & \csvcolii & \csvcoliii\%}
    \midrule
    \multirow{5}{*}{Malicious}%
    \begin{filecontents*}{data/causes-bad.csv}
cause,count,percent
Malicious behavior,8510,91.2013717715143
Unexpected behavior,564,6.044368234915872
JS parsing error,172,1.8433179723502304
Fixpoint not reached,78,0.8359232665309184
Extraction error,7,0.07501875468867217
\end{filecontents*}
\csvreader[late after line=\\]{data/causes-bad.csv}{}{& \csvcoli & \csvcolii & \csvcoliii\%}
    \bottomrule
  \end{tabular}
  \label{tab:failures-malware}
\end{table}

First, because \ourtool is designed for high recall, we manually investigated the 7 false negatives reported in \Cref{tab:comparison}
and confirmed that \emph{all} of them stem from documents that contain \emph{benign} JavaScript only. 
Specifically, 4 PDFs trigger a JavaScript \mintinline{javascript}{alert} message that encourages the user to open a malicious file
attachment, 2 PDFs contain benign form manipulation code only, and 1 PDF contains a malicious payload
that is embedded in a JavaScript function that is never called. 

Next, we investigated the causes of malware reports in the benign and
malicious benchmark sets to learn what causes the analysis to classify a sample as
malware. \Cref{tab:failures-malware} breaks down the causes and shows their
proportions relative to all malware reports in each of the two sets.

``Malicious behavior'' refers to a whitelist violation, \eg creation of a large
object, or a call to a vulnerable API method or to
\mintinline{JavaScript}{eval}.
As expected, it is the top cause for identifying \emph{true positives} in the
``Malicious'' benchmark set.
In the ``Benign'' set, ``Malicious behavior'' indicates the use of a
(historically) vulnerable API in a non-malicious way and such cases are
responsible for approximately a third of \emph{false positive} reports.
We plan to address some of these false reports in the next version of \ourtool
by inspecting the abstract values that reach vulnerable APIs. \Ie we will not report
PDFs where abstract interpretation can prove the absence of a malicious payload 
reaching a vulnerable API.

The remaining causes for malware reports, are instances of \ourtool
being conservative.
In both benchmark sets, the majority of conservative reports can be attributed to broken
JavaScript code (``\unexpected'' and JS parsing error) 
and PDF documents broken beyond repair (``Extraction error'').
Looking into the benign documents for which extraction fails, we found that the
majority (57) come from the hard-to-extract set, of which 13 were broken during
HTTP download (before being submitted to VirusTotal), and only two can be
opened with Acrobat Reader DC 2018.011.
A small fraction of malware reports is caused by
abstract interpretation aborting due to lack of precision
or timing out (``Fixpoint not reached'').
In all these cases, we cannot safely over-approximate the behavior of JavaScript code
and thus have to classify it as malicious.

``\Unexpected'' refers to cases where \ourtool encounters semantically
incorrect JavaScript code, such as property loads from undefined variables, calls to
undefined functions, and other behavior that should not occur in a
\emph{functionally correct} PDF document.
\begin{listing}[t]
\begin{minted}{JavaScript}
qwe = ('nhsthn','ntrht').substr;
var g = qwe();
\end{minted}
    \caption{Example of \unexpected exhibited by malware (excerpt)}
    \label{lst:lookup-example}
\end{listing}

However, we cannot simply conclude that these are honest mistakes (bugs) that can
safely be ignored, especially not in the context of JavaScript execution, where
engines are known for their lenient and non-standard ways of handling errors.
Indeed, obvious bugs also occur in malicious code:
\Cref{lst:lookup-example} shows an example of malware being
identified through \unexpected before its actual malicious behavior is
detected.
At line 1, the comma operator evaluates each of its operands (from left to right) and 
returns the value of the last operand \cite{mozilla-spec}. Then, \mintinline{JavaScript}{qwe} 
is assigned the \mintinline{JavaScript}{substr} function, which is not bound to any 
receiver object at this point in the program. Next, \mintinline{JavaScript}{qwe} is called at line 2 without a receiver 
object, resulting in an exception being thrown. \ourtool correctly identifies the call 
\mintinline{JavaScript}{qwe()} as faulty, and conservatively reports the input as malware.
We confirmed that this code snippet raises an exception in recent JavaScript engines 
(\eg Node.js v8.1.4).

After manual inspection, we concluded that the high number of ``\unexpected''
in the benign benchmark set is due to semantically incorrect code that either has
not been properly tested or maintained or even ended up inside a PDF
document by accident (\eg code that only makes sense in the context of a website).
As described in \Cref{sec:ExpSetupSub}, we handle one frequently occuring instance
falling into the last cateogry, but have not addressed all issues of this kind
due to time constraints.
The existence of such erroneous code can be explained by PDF viewers silently
ignoring JavaScript errors.

\subsection{Threat model and possible attacks}

In this section, we present a realistic threat model and explore potential
attacks against \ourtool.

We assume an attacker is trying to have malicious JavaScript code inside a PDF, originally reported as such
by \ourtool, misclassified as benign. The attacker can manipulate
the PDF document however they want. We further assume that the attacker has 
black-box access to \ourtool and can observe the outcome 
report only (\eg benign or malicious). We also assume that the attacker
can submit an unlimited number of PDFs to \ourtool. We further
assume that the attacker knows that \ourtool uses abstract
interpretation and a model of the PDF viewer runtime environment, but has
no access to it. Because we are interested in attacks against \ourtool
and not the JavaScript code extractor, we finally assume that all the JavaScript
code in the PDF can be extracted without error.

First, because \ourtool analyzes JavaScript code only, it is 
oblivious to mimicry attacks~\cite{Maiorca2013,Srndic2014} that alter the 
\emph{structure} of the PDF document without altering the malicious \emph{behavior}
of its payload. In~\cite{xu2016}, the authors showed how they could use genetic algorithms
to automatically alter the structure of 500 malware samples so that they successfully evade 
\toolname{Hidost}~\cite{Vsrndic2016} and \toolname{PDFRate}~\cite{Smutz2012}.

On the other hand, \ourtool
might be vulnerable to the following four threats: (1) zero-day vulnerabilities that
only exhibit whitelisted behaviors;
(2) discrepancies between the concrete semantics of a PDF viewer's JavaScript engine 
and the abstract semantics of the \toolname{SAFE} abstract interpreter; (3)
unsound approximations in the PDF-JS model; and (4) novel parser confusion attacks. 
In the following paragraphs, we explore each threat in more detail.
 
Obviously, zero-day vulnerabilities that are not exploited using JavaScript
are out of scope for \ourtool. Otherwise, for \ourtool
to miss a JavaScript-based exploit, the exploit must use a function that has 
been whitelisted. Indeed, if the vulnerability exploits
any function that is not whitelisted, \ourtool 
will interpret the function call as ``unexpected behavior'', and
conservatively report the PDF as malware. Hence, to exploit a zero-day vulnerability,
the attacker has to find a vulnerability in a whitelisted function, \emph{and} ensure
that the payload that exploits the vulnerability does not exhibit any known malicious 
behavior (\eg heap spraying). While not impossible, this attack 
is very unlikely.

The abstract semantics used in \ourtool is currently based on the 
ECMAScript 5.1 language specification and thus might differ from the ECMAScript version
 supported in PDF viewers. However, exploiting discrepancies between the abstract
and concrete semantics is far from trivial. First, using language features that 
are not supported by the abstract semantics of \ourtool will result in 
undefined behavior and the PDF being reported as malware. Therefore, the attacker 
must identify semantic discrepancies that are due to either implementation bugs 
in the viewer or in \ourtool and exploit them. This type of attack is also 
very unlikely, but not unheard of~\cite{takata2017}.
 
Because \ourtool relies on the analysis being conservative,
it makes it vulnerable to unsoundness bugs in the PDF-JS model. Indeed,
we assume that \ourtool strictly over-approximates all possible 
runtime execution paths, and every unsoundness bug in the PDF-JS
model introduces an opportunity for an attacker to hide malicious behavior behind
an under-approximation. Because the PDF-JS model was built manually, it is
prone to human errors and makes \ourtool vulnerable to unsoundness bugs.
A superior, but far more costly solution would be to automatically infer the 
PDF-JS model based on a set of concrete PDF viewer applications.

Finally, because the PDF specification is very complex, yet vague, and PDF
viewers often deviate from it, an attacker could try to trigger new parser
confusion attacks. However, because \ourtool treats extraction errors
\emph{conservatively}, and reports malware on extraction failures, an attacker 
would have to find a way to hide his payload from the extractor without causing
an extraction failure. While such an attack would require a dedicated
attacker, we believe it is the most serious threat to \ourtool.

\section{Related Work}

In this section, we give a brief overview of existing techniques for PDF
malware detection (a detailed survey and taxonomy can be found in
\cite{Nissim2015}) and explain how they compare to our analysis approach.
We also discuss semantic malware detection and JavaScript static analysis,
which are related fields of research.

\subsection{Static PDF Malware Detection}
\label{sec:rel:nonjs}
The first group of approaches proposed in academic literature that we consider as related work
analyzes a PDF document as a whole and does not analyze any embedded JavaScript code.
These techniques are categorized as \emph{Metadata Analysis} in~\cite{Nissim2015}.
Three properties of PDF documents are being used to derive a \emph{fingerprint}: 
document structure, metadata fields, and document content.
The related approaches \cite{Smutz2012,Maiorca2012,Nissim2014,Nissim2016,Vsrndic2016} rely on 
a set of known malicious PDF documents as training data to identify documents with a similar 
fingerprint as malware.

\toolname{Caradoc}~\cite{GuillaumeEndignoux2016} is an exception to the above. Endignoux et al.
focus on weaknesses in the PDF standard related to document structure. These
can be exploited to attack the parser implementation of a PDF viewer, \eg
to achieve a denial-of-service attack.

\cmpToUs %
When aiming to identify malicious PDF documents that exploit vulnerabilities in
the JavaScript runtime of a PDF viewer, our approach is more powerful,
since it does not depend on a training set and is not susceptible to mimicry attacks.

\subsection{Static PDF-JavaScript Malware Detection}
Similar to \ourtool, the related work in this section performs a static 
analysis of JavaScript code embedded in PDF documents.
However, unlike us, but similar to the approaches in \Cref{sec:rel:nonjs}, they identify
malicious JavaScript based on its similarity to known malicious samples.

\toolname{PJScan}~\cite{Laskov2011} and Vatamanu's approach~\cite{Vatamanu2012} both
perform lexical analysis of the extracted JavaScript code,
and use machine learning techniques
to classify the code as malicious or non-malicious.
In~\cite{Karademir2013}, the authors describe of use \toolname{NiCad}, an
existing tool for detecting code clones, for the same purpose.

\cmpToUs %
The approaches above are always less powerful than our static analysis, because
they restrict themselves to lexical analysis of JavaScript code and do not take
its semantics into account.
Both approaches rely on the similarity of (possibly obfuscated) JavaScript code
to known malicious code, and might be defeated by novel obfuscation patterns.

\subsection{Dynamic PDF Malware Detection}
All approaches in this section rely on the execution of PDF-embedded JavaScript
code, either in its native or synthetic runtime environment, for analysis of
its behavior.

\toolname{MDScan}~\cite{Tzermias2011} and \toolname{PDF Scrutinizer} execute
the extracted JavaScript code in a synthetic environment, and aim to detect the
presence of malicious payload (so called \emph{shell code}) in the execution
state (as part of strings and variables). \toolname{PDF Scrutinizer} applies
further heuristics to identify execution patterns typical of malicious code.
Other uses of dynamic analysis proposed in literature do not compare directly 
to our approach because they are not suitable as stand-alone analysis tools.
For example, \toolname{MPScan}~\cite{Lu2013} and
\toolname{FCScan}~\cite{schade2013fcscan} propose to integrate with the PDF
viewer software, and \toolname{ShellOS}~\cite{Snow2011} integrates with the
underlying operating system to detect attacks (on-the-fly) during runtime.

\cmpToUs
Dynamic analysis techniques are prone to miss feasible program behavior,
because actual execution depends on inputs and the execution environment.
In the context of malware analysis, the malware author can actively target a
specific environment, thus preventing the detection of malicious behavior in the
analysis environment.
For example, the de-obfuscation of exploit code might be
triggered only in a specific target environment. This undermines the dynamic
analysis-based detection techniques described above.
Furthermore, unlike our static analysis-based approach, none of the existing dynamic analyses
tries to exhaustively explore all possible behavior of the JavaScript code.

\subsection{Semantic-Based Malware Detection}
Semantic approaches use techniques from program analysis and formal methods to
lift malware detection from syntactic features to the level of program
semantics.
For example, semantic malware detectors use
theorem proving~\cite{Christodorescu2005} or model checking~\cite{Kinder2005}
to match a program, based on its semantic properties (\eg instruction sequences),
against a template derived from actual malware.
In general, these approaches are more powerful than signature matching, but
still prone to evasion by obfuscation.
Preda et al.~\cite{Preda2008} introduce a theoretical framework for semantic malware
detection using abstract interpretation.
It assumes the availability of \emph{perfect oracles}, which return perfect
information related to a program's semantic properties or behavior (\eg its
exact control flow), and shows that whether a detector can overcome a
particular obfuscation, depends on the chosen abstract semantics, \ie the right
level of abstraction.
More recently, \cite{Palahan2013} uses statistical analysis of program behavior
recorded during dynamic analysis to identify malware. This avoids the difficulty
of static reasoning, but introduces the possibility of dynamic analysis missing
malicious behavior.

\cmpToUs
Our work can be seen as an instance of malware detection based on Preda's \emph{interesting
actions}~\cite[Section~5]{Preda2008}. However, we use a policy to whitelist acceptable behavior
and thus do not have to rely on actual malware as templates for malicious
behavior.
Furthermore, our conservative strategy enables us to have a practical malware detector
in the absence of "perfect oracles"

\subsection{Web JavaScript Malware Detection}
The primary vector for JavaScript-based malware remains web browsers. Prominent work
in this area includes Nozzle~\cite{ratanaworabhan2009}, Zozzle~\cite{curtsinger2011}, and 
Rozzle~\cite{kolbitsch2012}. Nozzle is a dynamic, in-browser approach that uses heap 
sampling to detect heap-spraying attacks. Similar to other dynamic approaches, Nozzle
requires an instrumented environment, a browser in this case, and induces a performance
overhead. Zozzle is a mostly static approach that reduces the performance overhead
of Nozzle. It uses a Na{\"i}ve Bayes classifier, trained on syntactic features of the
JavaScript code, to classify programs as benign or malicious. Zozzle relies on the
browser's JavaScript interpreter to de-obfuscate the code before performing feature
extraction and classification. Because both Nozzle and Zozzle rely on an 
instrumented browser environment, both approaches are susceptible to miss malware that exhibits malicious
behavior on other environments only. Rozzle addresses this limitation through the use of
symbolic execution to emulate different runtime environments in a single instrumented browser.

\cmpToUs
Through the use of abstract interpretation, our work detects heap spraying (like Nozzle), 
performs de-obfuscation (like Zozzle), and simulates different runtime environments 
(like Rozzle) \emph{entirely} statically. Furthermore, \ourtool can statically
reason about the runtime behavior of the code, making it more powerful than syntax-based
approaches.

\subsection{JavaScript Static Analysis}
The dynamic nature of JavaScript and its lack of static guarantees make it a difficult target for static analysis.
This shortcoming is magnified by the inherent complexity of the most common
use of JavaScript, as client-side web application code running inside an equally complex
browser environment.
At this point in time, state-of-the-art tools
based on dataflow analysis~\cite{wala} or precise abstract interpretation~\cite{tajs,safe} 
can successfully analyze libraries and small applications, but do not scale to
real-world JavaScript code in general.

We are nonetheless successful in using the very same techniques to analyze
JavaScript in the context of malware detection.
This is due to two reasons: (1) JavaScript code embedded in PDF documents is
not as complex as code written for the web; (2) malware detection warrants a 
conservative strategy where unknown behaviors are interpreted as malicious.
\section{Conclusion}
We presented a novel approach for detecting malicious JavaScript embedded in PDF
documents that uses abstract interpretation---a static program analysis
technique---at its core.
Using the results of abstract interpretation in a conservative manner, our
malware detection is designed for and achieves very high recall. Furthermore,
with an average runtime of less than 4 seconds per document, we showed how 
traditionally ``heavy-weight'' abstract interpretation tools can be used in
practice, given the right abstraction (e.g. the PDF-JS model).
By addressing all known limitations of existing PDF JavaScript code extractors, we
showed that PDF malware detectors that analyze the embedded JavaScript code can be
used in practice. We also showed how \ourtool resists obfuscation, parser confusion, 
and mimicry evasion attacks that subvert existing malware detector tools.
Finally, through comprehensive experimental evaluation, we have shown that our approach
achieves almost perfect recall, and comparable precision to state-of-the-art tools.

\section*{Acknowledgment}

The authors would like to thank Phil Boutros and Joe Keslin from the Oracle
Clean Content team for their support.

\bibliographystyle{plain}

\end{document}